\documentclass[11pt]{article}
\pdfoutput=1

\usepackage[margin=1in, paperwidth=8.5in, paperheight=11in]{geometry}

\usepackage{graphicx}
\usepackage{float}
\usepackage{caption}
\usepackage{subcaption}
\usepackage{fp}
\usepackage{pgfplots}
\pgfplotsset{compat=1.13}

\usepackage{tikz}

\usepackage{verbatim}

\usepackage{booktabs}

\usepackage[numbers]{natbib}
\usepackage{url}

\begin{document}

\title{ Extending the D-Wave with support for Higher Precision Coefficients }
\author{John E. Dorband\\
	Department of Computer Science and Electrical Engineering\\
	University of Maryland, Baltimore County\\
	Maryland, USA\\
	\texttt{dorband@umbc.edu}}
\date{\today}
\maketitle

\begin{abstract}

D-Wave only guarantees to support coefficients with 4 to 5 bits of resolution or precision.
This paper describes a method to extend the functionality of the D-Wave to solve problems that
require the support of higher precision coefficients.

\end{abstract}

\section{Introduction}\label{sec:intro}

The D-Wave\citep{Dwave13} is an adiabatic quantum computer\citep{Farhi00,Giuseppe08} which supports the
following objective function:
\begin{equation}\label{eq:obfunc}
F = {\sum\limits_i a_i q_i + \sum\limits_i \sum\limits_j b_{ij} q_i q_j}
\end{equation}
where $q_i\in\{-1,1\}$ are the qubit values returned by the D-Wave, and $a_i\in[-2,2]$ and $b_{ij}\in[-1,1]$ are 
the coefficients given to the D-Wave associated with the qubits and the qubit couplers respectively. 
This however does not tell the whole story.

An algorithm has been developed, multi-qubit correction, MQC, that
reduces a set of D-Wave result samples to a single sample that has a objective function value that is less-than
or equal-to the objective function value of any sample of D-Wave result sample set.
This algorithm presumes that the D-Wave result samples contain groups of qubits that can be used to construct
a more optimal solution to the objective function. 
This has borne out to be true, as has been presented in the paper \citep{Dorband17}.

This presumption that D-Wave result samples contain information that can be utilized to construct a better solution
to the objective function is utilized here to extract information from multiple variations of the objection
function to find an optimal solution to an objective function that is based on coefficients of greater precision.

The precision of the coefficient values that are passed to the D-Wave hardware influences the 
values of the qubits that are returned by the D-Wave hardware.
Currently, D-Wave only guarantees to take advantage of 4 to 5 bits precision of 
the floating point coefficient values passed to it by the user.
This paper presents an algorithm, HPE (High Precision Enhancement), that will allow the D-Wave to in effect support higher 
precision coefficient values.
This means that more bits of the values passed to it will be effectively utilized.

Precision here is measured in terms of bits of resolution. For example, 4 bit resolution represents the ability to
represent the numbers 0 to 15 in integer increments, or the values from zero to one in increments of $\frac{1}{16}$. 
In general, n bit resolution can represent the values between 0 and 1 in increments of $\frac{1}{2^n}$.
And since D-Wave values can be positive or negative, and additional bit is need as a sign bit.
Thus, $n+1$ bit resolution represents the values between 1 and -1 in increments of $\frac{1}{2^n}$.

HPE constructs multiple version of the objective function, all of which have the same minima and maxima.
This is done by multiplying the objective functions by different constants (Eq. \ref{eq:cons_obfunc}).
\begin{equation}\label{eq:cons_obfunc}
F^k = {\sum\limits_i c_k a_i q_i + \sum\limits_i \sum\limits_j c_k b_{ij} q_i q_j}
\end{equation}
where $k$ is the objective function version and $c_k$ is the version constant.
Note that if $a_i$ and $b_{ij}$ are multiplied by the same constants $c_k$ the minima of $F$ and $F^k$ 
are the same.
Thus D-Wave should return a valid set of $q_i$ for either $F$ and $F^k$.
We are assuming here that $a_i$ and $b_{ij}$ are floating point numbers and that the D-Wave support coefficients
with precision up that of double precision floating number.
This is not however the case as has already been pointed out.
And arguably, $F$ and $F^k$ are not the same once the coefficients $a_i$ and $b_{ij}$ are truncated to a range 
of values $[-2,2]$ and $[-1,1]$ respectively and are reduce to the precision supported by the hardware. 

So in reality each $F^k$ represent a different function to the D-Wave hardware.
The previously stated presumption is used here, that D-Wave result samples, though they are not the global 
solution, may contain information that lead to the global minimum.
For the purpose of this algorithm the presumption is that the different versions of $F^k$ will return sample sets
that contain information that lead to a better solution, if not the global solution, of the given 
higher precision objective function $F$.

\section{The Higher Precision Enhancement Algorithm (HPE)}\label{sec:hpe}

\subsection{The Algorithm}\label{sec:method}

For each function version $F^k$ a sample set $S^k$ is obtained from the D-Wave.
Each sample set may contain anywhere from 1000 to 10,000 samples, $s^k_i$.
Each sample $s^k_i$ is a set of qubit values, $q_j$, that represent a solution to $F^k$.

HPE extensively utilizes MQC to reduce sample sets to single samples.
The algorithm starts by taking a sample, $s^k_i$, from each samples set, $S^k$, to make a sample set, $T^h$.
Then MQC is used to reduce $T^h$ to a single sample $t^h$.
These samples, $t^h$, are formed into a sample set $H$. 
And finally, MQC is used to reduce $H$ to a single sample, $h$, which is HPE's solution for $F$.

Although the coefficients of $F^k$ are used to provide coefficients to the D-Wave to obtain result samples,
MQC always uses the coefficients of $F$ to reduce the sample sets.

\subsection{The Scaling}\label{sec:scale}

HPE is simple, but the choice of the scaling parameters are critical 
to the success of the algorithm, but may not be difficult to determine.

The common scaling factor, $d$, is the maximum absolute value of magnitude of 
the minimums and maximums of the $a_i$ and $b_{ij}$.
The initial scaling factor, $c_0$, is a multiple of the inverse of $d$, so as to allow the 
resolution of the higher resolution of larger value coefficients.
The choice of $c_k$ is very much dependent on the greater precision desired as well as the dynamic range of
the coefficients.

In the following test cases $c_0=\frac{1}{8 d}$ and $c_{k+1}=c_{k} \sqrt{2}=c_{0} (\sqrt{2})^k$
(i.e. $F^k$ scales by a half a bit of precision as k is incremented by 1).
In the following tests, HPE used 20 versions of $F^k$ where $k$ varies from 0 to 19.

\section{The Test Results}\label{sec:tests}

The tests are based on pseudo-random number generated coefficients, values of $a_i$ and $b_{ij}$.
These coefficients are quantized to the level of precision required for the test.
A ground truth coefficient set for $F$ is generated at a specified precision, then 
a series of coefficient sets for $F^k$ are generated at a lower precision.
The ground truth set for $F$ represents the higher precision problem to be solved.
While the lower precision sets $F^k$ represent a set of problems which are solved by 
lower precision quantum hardware.
This is done to allow the comparison of how well a high precision problem can be solved 
with lower precision hardware.

The objective function, $F$, is defined as the base or anchor problem, the problem that defines 
mathematical ground truth of the problem for which the minimum is to be found, the solution.
The base problem consists of both the coefficients and the precision to whick the coefficients are
to be quantized.
The solution sample returned by the D-Wave with the minimum value for $F$ is defined as $dw$.
The solution sample obtained from MQC applied to the samples return by the D-Wave for $F$ is defined as $m$.
The solution sample obtained from HPE when applied to the samples returned by the D-Wave from
solving all of $F^k$ objective functions is $h$.

The feasibility tests(section \ref{sec:FeasTest}) is to test the feasibility of solving high 
precision problems with low precision hardware.
The feasibility tests simulates very low precision virtual-hardware by
creating very low resolution quantization of $F^k$.
The base problem $F$ is quantized at a resolution that represents the level that the D-Wave should be able
to support.

The high precision test(section \ref{sec:HPTest}) is intented to show how the D-Wave results alone or the D-Wave
results post-processed with MQC compares with HPE utilizing the D-Wave.
The high precision test is to test problems with a much higher 
precision requirement than the D-Wave supports, while HPE utilizes the D-Wave to solve the problem with 
$F^k$ solved on the D-Wave's lower precision hardware.

And finally, the unconstrained test(section \ref{sec:UnContTest}) neither quantizes the 
problem coefficients or the HPE solver coefficients, 
(i.e. all coefficients are unmodified double precision numbers).

Note that, even though D-Wave only claims that the D-Wave hardware only supports 4 to 5 bit precision,
the D-Wave is assumed in this paper to optimistically support 9 bit precision.

In Tables \ref{table:threebitHardware} thru \ref{table:DWaveHardware}, 
'BP' is the base/anchor problem precision, 
'HP' is the hardware precision (real or pseudo hardware),
'Cases' is the number of difference sets of coefficients used to  define the problem to be solved,
'Samples' is the number of sets of qubit values requested from the D-Wave for each problem set of coefficients to be solved,
'$dw < h$' represents the number of cases where the raw D-Wave results have a lower value than the HPE result,
'$m < h$' represents the number of cases where the MQC result value had a lower value than the HPE result,
'$m = h$' represents the number of cases where the MQC result value had the same value as the HPE result, and
'$h < m$' represents the number of cases where the HPE result value had a lower value than the MQC result.

In all tests case presented in this paper the MQC results had a lower value than the raw D-Wave results.
Due to the nature of the MQC algorithm MQC will always have as low or lower value than the raw D-Wave results.

\subsection{Feasibility Test}\label{sec:FeasTest}

These feasibility tests are the initial test to show that the algorithm successfully uses 
lower precision hardware to solve a higher precision problem.
For this test the ground truth precision is 9 bits, one sign bit and 8 bit resolution,
and the pseudo-quantum hardware machine precision is 3 bits, one sign bit and 2 bit resolution.

\begin{table}[h]
\begin{tabular}{|l|c|c|c|c|c|c|c|c|c|c|c|}
\hline
BP & HP & Cases & Samples & $dw < h$ & $m < h$ & $m = h$ & $h < m$       \\ \hline
9  & 3  & 1000 &  1000 & 9 & 775 & 106 & 119     \\ \hline
9  & 3  & 100  & 10000 & 1 & 56 & 28 & 16        \\ \hline
17 & 3  & 100  & 10000 & 0 & 59 & 30 & 11        \\ \hline
25 & 3  & 100  & 10000 & 0 & 56 & 34 & 10        \\ \hline
33 & 3  & 100  & 10000 & 0 & 52 & 38 & 10        \\ \hline
\end{tabular}
\centering
\caption{Various Precision Problems Running on 3 Bit Precision Pseudo-hardware.}
\label{table:threebitHardware}
\end{table}

Table \ref{table:threebitHardware} shows the results of the feasibility tests.
If the base or ground truth problem is of order the precision of the actual D-Wave,
the D-Wave does well. In some cases it did do better than HPE though never better than MQC.
In most cases however HPE does better than the D-Wave alone even though it was working with far less precision.
Arguably, the D-Wave with MQC post processing did do better in most cases than HPE with much lower precision.
HPE with the lower precision did, in some cases, better than the D-Wave with MQC post processing.
This shows that it is feasible to obtain a solution of a high precision problem with much lower
precision hardware.

\subsection{High Precision Test}\label{sec:HPTest}

The purpose of the high precision test is to show how well HPE performs using the D-Wave over the 
D-Wave with out HPE processing.
Table \ref{table:eightbitHardware} shows the results of the high precision tests.

\begin{table}[h]
\begin{tabular}{|l|c|c|c|c|c|c|c|c|c|c|c|}
\hline
BP & HP & Cases & Samples & $dw < h$ & $m < h$ & $m = h$ & $h < m$       \\ \hline
9  & 9  & 100  & 1000 & 0 & 0 & 37 & 63       \\ \hline
17 & 9  & 100  & 1000 & 0 & 0 & 48 & 52       \\ \hline
25 & 9  & 100  & 1000 & 0 & 0 & 45 & 55       \\ \hline
33 & 9  & 100  & 1000 & 0 & 0 & 45 & 55       \\ \hline
41 & 9  & 100  & 1000 & 0 & 0 & 44 & 56       \\ \hline
49 & 9  & 100  & 1000 & 0 & 0 & 48 & 52       \\ \hline
\end{tabular}
\centering
\caption{Various Precision Problems Running on 9 Bit Precision Hardware.}
\label{table:eightbitHardware}
\end{table}

In no case was the D-Wave alone or with MQC better than HPE and in more than half the cases
HPE was better than the D-Wave with MQC, otherwise it was the same as MQC.

\subsection{Unconstrained Test}\label{sec:UnContTest}

Table \ref{table:DWaveHardware} shows the results of the unconstrained tests.
In the unconstrained test, neither the coefficients of $F$ or $F^k$ are preprocessed by quantization.
All procession is by sending non-quantized double precision floating point coefficients to be
solved by only D-Wave processing. 
The results vary little from the high precision tests.
'dbl' refers to full double precision floating point.

\begin{table}[h]
\begin{tabular}{|l|c|c|c|c|c|c|c|c|c|c|c|}
\hline
BP & HP & Cases & Samples & $dw < h$ & $m < h$ & $m = h$ & $h < m$       \\ \hline
dbl & dbl  & 1000  & 1000 & 0 & 0 & 39 & 61      \\ \hline
\end{tabular}
\centering
\caption{49 bit Precision Problem Running on Unconstrained D-Wave Hardware.}
\label{table:DWaveHardware}
\end{table}

\section{Conclusion}\label{sec:Conclusion}

HPE (high precision enhancement) is a simple algorithm for processing objective functions which require
higher precision coefficient than the current or possibly any future quantum hardware will be able to support.
The feasibility test results indicate that if all you need is the precision that the hardware supports,
then quantum plus MQC is possibly sufficient.
However if the support of precision greater than that supported by the hardware is needed, then HPE may give a
significant edge over quantum hardaware with MQC post-processing.

\section*{Acknowledgment}

The author would like to thank Michael Little and Marjorie Cole of the NASA Advanced Information Systems 
Technology Office for their continued support for this research effort under grant NNH16ZDA001N-AIST16-0091 
and to the NASA Ames Research Center for providing access to the D-Wave quantum annealing computer. 
In addition, the author thanks the NSF funded Center for Hybrid Multicore Productivity Research and 
D-Wave Systems for their support and access to their computational resources.  

\bibliography{QuantumBib}{}

\begin{thebibliography}{4}
\providecommand{\natexlab}[1]{#1}
\providecommand{\url}[1]{\texttt{#1}}
\expandafter\ifx\csname urlstyle\endcsname\relax
  \providecommand{\doi}[1]{doi: #1}\else
  \providecommand{\doi}{doi: \begingroup \urlstyle{rm}\Url}\fi

\bibitem[{Dorband}(2017)]{Dorband17}
J.~E. {Dorband}.
\newblock {Improving the Accuracy of an Adiabatic Quantum Computer}.
\newblock \emph{ArXiv e-prints}, May 2017.

\bibitem[{Farhi} et~al.(2000){Farhi}, {Goldstone}, {Gutmann}, and
  {Sipser}]{Farhi00}
E.~{Farhi}, J.~{Goldstone}, S.~{Gutmann}, and M.~{Sipser}.
\newblock {Quantum Computation by Adiabatic Evolution}.
\newblock \emph{eprint arXiv:quant-ph/0001106}, January 2000.

\bibitem[Inc.(2013)]{Dwave13}
D-Wave~Systems Inc.
\newblock {The D-Wave 2X Quantum Computer: Technology Overview}.
\newblock \url{http://www.dwavesys.com/resources/publications}, 2013.
\newblock [Online; accessed 18-May-2016].

\bibitem[{Santoro} and {Tosatti}(2008)]{Giuseppe08}
Giuseppe {Santoro} and Erio {Tosatti}.
\newblock Optimization using quantum mechanics: quantum annealing through
  adiabatic evolution.
\newblock \emph{Journal of Physics A: Mathematical and Theoretical},
  41\penalty0 (20):\penalty0 209801, 2008.
\newblock URL \url{http://stacks.iop.org/1751-8121/41/i=20/a=209801}.

\end{thebibliography}

\end{document}